\documentclass{blois}

\def\Journal#1#2#3#4{{#1} {\bf #2}, #3 (#4)}

\def\be{\begin{equation}}
\def\ee{\end{equation}}
\def\bea{\begin{eqnarray}}
\def\eea{\end{eqnarray}}

\newcommand{\Photo}{}

\begin{document}
\vspace*{4cm}
\title{NEUTRINOS FROM DENSE ENVIRONMENTS}

\author{M. CRISTINA VOLPE}

\address{Centre National pour la Recherche Scientifique, Université Paris Cité, \\ Astroparticule and Cosmology Laboratory,
F-75013 Paris, France}

\maketitle\abstracts{Neutrinos from dense environments are unique laboratories for astrophysics, particle physics and many-body physics. 
They tell us about the last stages of the gravitational core-collapse
and the explosion of massive stars. These elusive particles are also tightly linked to heavy elements synthesis in gravitational core-collapse supernovae and binary neutron star mergers, or play a pivotal role 
at the MeV epoch during the Universe expansion. We highlight theoretical and observational aspects of this interesting domain, in particular for the future measurement of neutrinos from the next core-collapse supernova, and of the diffuse supernova background, whose discovery
might lie in the forthcoming future.}

\section{Introduction}
\noindent
Dense environments are unique laboratories for astrophysics, particle physics and many-body physics. Neutrinos in such environments 
constitute a weakly interacting many-body system. They represent unique probes of the stellar interior, during
spectacular phenomena such as core-collapse supernova explosions, in the search for new physics, and impact the synthesis of elements in astrophysical environments. 

Noteworthy progress has been made in our description and understanding of how neutrinos evolve in dense environments and eventually change their flavor. Such theoretical developments have gone in parallel with key steps in the solution of two longstanding open issues in astrophysics, that have benefitted of unexpected exceptional observations, from SN1987A\cite{Kamiokande-II:1987idp,Bionta:1987qt,Alekseev:1988gp} and GW170817 \cite{LIGOScientific:2017ync}. These key questions are identifying the sites and conditions that produce elements heavier than iron in the rapid neutron-capture nucleosynthesis process (known as the $r$-process\cite{Cowan:2019pkx})
on the one hand, and unravelling the mechanism(s)  that make gravitational core-collapse supernovae\footnote{Supernovae are massive stars that, at the end of life, undergo either thermonuclear explosions -- SNe type Ia -- or gravitational core-collapse -- SNe types II and Ib/c. } explode on the other. Thus, understanding how neutrinos propagate and change flavor in dense environments is fundamental  theoretically and for future observations of core-collapse supernova neutrinos, of the diffuse supernova neutrino background and for identifying the site(s) where $r$-process elements are made (see\cite{Volpe:2023met} for a recent review). 

Observationally, GW170817 brought the first concomitant observation of gravitational waves from a binary neutron star merger, a short gamma-ray burst and a kilonova\cite{LIGOScientific:2017ync}, powered by the decay of radioactive elements. Among the wealth of information, that this event brought, is the confirmation of the (originally considered as very speculative) idea that the ejecta of a compact binary system could have $r$-process elements, as first suggested fifty years back by Lattimer and Schramm\cite{Lattimer:1974slx}.  With GW170817 we had the first evidence of the presence of $r$-process elements (e.g. lanthanides) in the ejecta of a binary neutron-star merger, while studies show that binary-neutron star mergers are likely the main site for the $r$-process, since astrophysical conditions are met to produce up to the heaviest elements (strong $r$-process). 

Heavy elements nucleosynthetic abundances depend on nuclear properties of a large ensemble of unstable nuclei, many of which cannot be measured in experiments on Earth. Neutrinos and neutrino-interactions also play an important role determining the neutron-to-proton ratio, a key nucleosynthetic parameter for the ejecta composition of binary neutron star mergers and of core-collapse supernovae. Furthermore numerous investigations have shown that neutrino flavor evolution is also likely to influence the heavy elements abundances in the $r$-process (see for example\cite{Volpe:2023met,Balantekin:2004ug,Duan:2010af,Wu:2017drk,George:2020veu,Xiong:2020ntn,Balantekin:2023ayx}). 

Since $r$-process abundances rely on nuclear properties, it is essential to evaluate the impact of nuclear uncertainties on the nucleosynthetic outcomes\cite{Barnes:2020nfi,Lund:2022bsr} (in particular from $\beta$-decay, fission and neutron-capture rates, or nuclear masses). In this respect, future measurements at the Facility for Rare Isotope Beams (FRIB)\cite{FRIB} should help reducing uncertainties. 
It is to be noted that neutrinos play an important role also on other nucleosynthesis processes\cite{Arcones:2022jer}, such as the $\nu$-p process\cite{Frohlich:2005ys} where neutrinos can produce proton-rich ejecta, and the recently identified\cite{Balantekin:2023ayx} "$\nu$i-process", where neutrino interactions help producing 
proton-rich low-mass and neutron-rich high-mass nuclei.

Well before  GW170817, on February 23rd, 1987, SN1987A\footnote{Koshiba received the 2002 Physics Nobel Prize (1/4) for this unique observation, with R. Davis (1/4) for the pioneering observation of solar neutrinos and R. Giacconi (1/2) for X-ray astronomy.} occurred in the Large Magellanic Cloud, when Sk-69$ ^\circ$202 exploded in this satellite galaxy of our Milky Way. It is the only event in which neutrinos from the gravitational core-collapse of a massive star were observed, and the first naked-eye sighting supernova, since Kepler's SN1604 (Ia).

SN1987A was unique in many respects, but particularly because we observed the neutrinos from the collapsed core. Only 24 $\bar{\nu}_e$ events were detected by the water Cherenkov detector Kamiokande-II (KII)\cite{Kamiokande-II:1987idp}, the Irvine-Michigan-Brookhaven (IMB)\cite{Bionta:1987qt} and Baksan Scintillator Telescope  (BST)\cite{Alekseev:1988gp}.  Five hours before other observations the Mont Blanc Liquid Scintillator Detector (LSD)\cite{Aglietta:1987it} recorded 5 events that keep being controversial.   A Bayesian analysis\cite{Loredo:2001rx} and subsequent analysis\cite{Ianni:2009bd} of the time signal favored the presence of an accretion+cooling phase in concordance with the delayed neutrino-mechanism of Bethe and Wilson\cite{Bethe:1985sox}. On the other hand 2D-likelihood analyses agreed on the expected average neutrino energies and total luminosities. Recently analyses confirmed previous findings\cite{Ivanez-Ballesteros:2023lqa,DedinNeto:2023hhp}. 
Interestingly, after more than thirty years searches, SN1987A remnant has now been identified:  it is a neutron star thermally obscured by dust\cite{Alp:2018oek,Cigan:2019shp,Page:2020gsx}. Detailed supernova (one-dimensional) simulations that fit reasonably well SN1987A observations\cite{Fiorillo:2023frv} became available recently. Since their observation, SN1987A neutrino events have been a unique laboratory for astrophysics, particle physics and the search for new physics. 

Among the unknown neutrino properties there is neutrino non-radiative two-body decay producing  a massless scalar or pseudo-scalar boson. An example is the
Majoron\cite{Chikashige:1980ui} associated with total lepton-number breaking. 
Figure 1 (left) shows the sensitivity to the lifetime-to-mass ratio $\tau/m$ of the decaying eigenstates for neutrino non-radiative decay in vacuum, considering either Earth-based experiments, or astrophysical neutrino sources.  Due to neutrino decay the neutrino flux depletes exponentially, with exponent (- L/E $\times m/\tau$), with L the distance between the source and the detector, and E the neutrino energy.  As a consequence, astrophysical sources such as the Sun, core-collapse supernovae or the diffuse supernova neutrino background\footnote{This is the neutrino background 
produced by past supernova explosions.} (DSNB) have a unique sensitivity for the largest $\tau/m$ values. Note that cosmology is also informative\cite{Chen:2022idm} on the large lifetime-to-mass ratios.

Based on a 3$\nu$ framework, a 7D-likelihood analysis of the 24 neutrino 
events measured by KII, IMB and BST (see Figure 1, right) provided the limit for neutrino non-radiative two-body decay\cite{Ivanez-Ballesteros:2023lqa} of 
$\tau/m > 2.4~(1.2)~\times~10^5$ s/eV at 68$\% ~(90~ \%$) C.L. (lifetime-to-mass ratio for the mass eigenstates $\nu_2$ and $\nu_1$, inverted mass ordering). This lower bound is tighter than those obtained by Earth-based experiments and is competitive with the model-dependent limits from cosmology. Moreover, competitive bounds on neutrino-Majoron couplings were also obtained thanks to the first likelihood analysis based on the SN1987A events and the spectral distortion induced by neutrino-decay in matter\cite{Ivanez-Ballesteros:2024nws}.   

\begin{figure*}[t]
    \centering
    \includegraphics[scale=0.47]{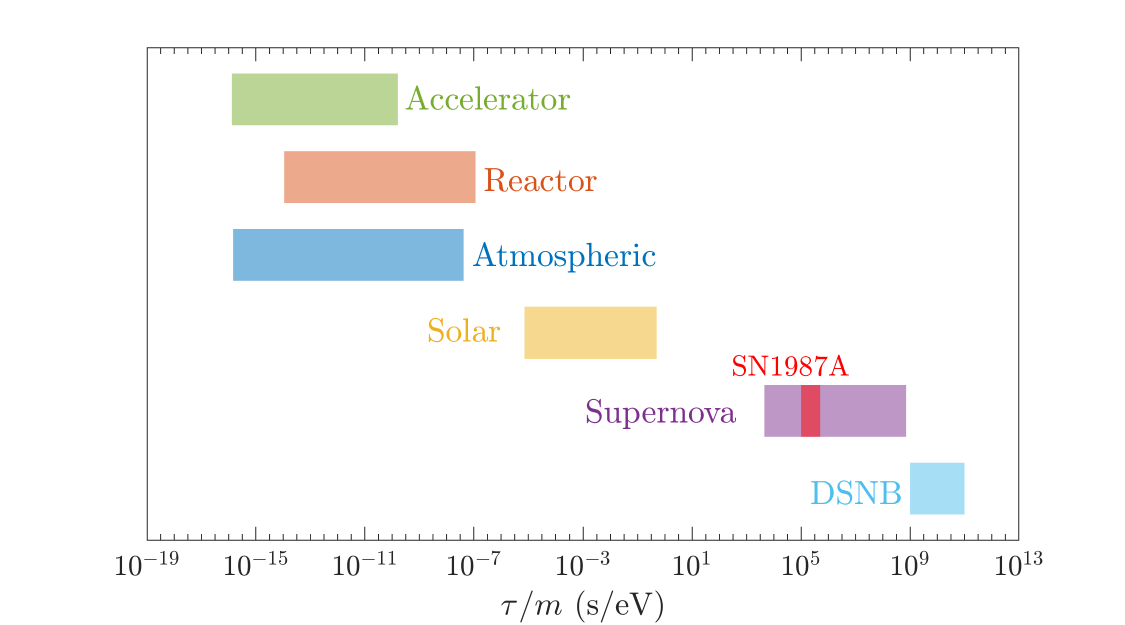}
    \includegraphics[scale=0.45]{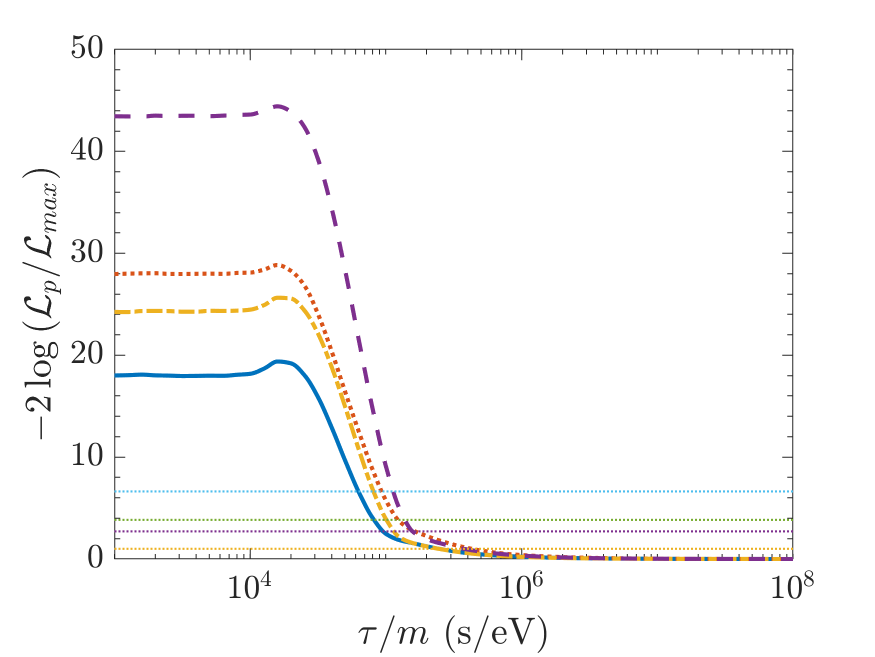}
    \caption{Neutrino non-radiative decay in vacuum. Left figure:
   Typical lifetime-to-mass ratios for neutrino non-radiative decay for different neutrino sources. The bands are obtained by considering typical distances and neutrino energies for the corresponding experiments, and considering either almost full decay ($1 \%$ of the initial flux) or practically no decay ($99 \%$ of the initial flux). Also the band corresponding to SN1987A is shown, corresponding to $E  \in [10, 50]$ MeV. 
    Right figure: Profile likelihood ratios from the 7D likelihood analysis of SN1987A events from KII, IMB and BST, for the case of
     inverted mass ordering.  The calculations of the neutrino fluxes include flavor conversion due to the Mikheev-Smirnov-Wolfenstein$^{32,33}$ (MSW) effect and neutrino decay in a 3$\nu$ flavor framework. 
The dashed straight lines from bottom to top correspond
to 68 $\%$, 90 $\%$, 95 $\%$, and 99 $\%$ CL.
The curves correspond to results:  i) without BST events and with background (full line);
ii) with BST events and background (dot-dashed line), iii) without BST events and without background (dotted)
and with BST events and without background (dashed line)$^{30}$.}
    \label{fig:decay}
\end{figure*}

As for the explosion mechanism(s) of very massive stars, the six-decade quest has gone through a crucial step forward every decade.  Since a decade an emerging consensus in the supernova community that most supernovae explode due to the delayed neutrino-heating mechanism. In this scenario, first suggested by Bethe and Wilson\cite{Bethe:1985sox},  neutrinos efficiently reheat the shock aided by neutrino driven convection, turbulence and hydrodynamic instabilities (the standing-accretion-shock-instability or SASI)\cite{Mezzacappa:2020pkk}. Note however that the role of SASI for all progenitors is still debated. If a new supernova explodes then the SASI would leave characteristic imprints
on the neutrino signal, as pointed out by several studies\cite{Tamborra:2014hga,Walk:2019ier,Walk:2019miz}. Rotation and magnetic fields can also help explosions, and their effects are being studied\cite{Kotake:2005zn,Kuroda:2020bdq}. It is to be noted that more than a decade ago, a neutrino-hydrodynamical instability
breaking the spherical symmetry was also uncovered: the lepton-number emission self-sustained asymmetry\cite{Tamborra:2014aua} (LESA), which shows a pronounced
$\nu_e$ and $\bar{\nu}_e$ dipole pattern in the supernova neutrino emission. 

\section{Theoretical aspects}
Since almost two decades our understanding on how neutrinos evolve and change flavor in dense environments has undergone profound developments, that are also crucial for our predictions for future observations\cite{Volpe:2023met,Volpe:2015rla,Duan:2010bg,Mirizzi:2015eza,Tamborra:2020cul,Horiuchi:2018ofe}. In dense environments, the neutrino density can become comparable
and even exceed the matter density, as in supernovae that undergo gravitational core-collapse, or binary compact merger remnants.
In a gravitational core-collapse supernova, matter densities reach from $10^{10}$ g/cm$^3$ to a few $10^{14}$ g/cm$^3$ in the central regions, the limit of matter compressibility,
and produce large neutrino fluxes with up to $10^{57}$-$10^{58}$ neutrinos with tens of MeV.

An interesting consequence of having large neutrino fluxes is that the neutral current neutrino-neutrino interaction becomes important, as Pantaleone first pointed out\cite{Pantaleone:1992eq} in the supernova context more than thirty years ago.
Besides sizable neutrino-neutrino interactions, during the explosion of a gravitational core-collapse supernova, neutrino change their flavor due to 
shock wave effects and turbulence.  

Already more than ten years ago the mean-field approximation used to determine how neutrinos change flavor in dense media was questioned.
First corrections beyond the mean-field were discussed, coming from corrections to the saddle-point approximation in a path integral approach\cite{Balantekin:2006tg}. Then,
using the Born-Bogoliubov-Green-Kirkwood-Yvon hierarchy, contributions from the pairing and the helicity correlators were also pointed out\cite{Volpe:2013uxl}.    
Solving the neutrino quantum kinetic equations is a 7-dimensional problem, since the Liouville operator depends on time, the three spatial coordinates and momenta.
As such, it has been fully solved recently only in the Early Universe context\cite{Froustey:2020mcq,Bennett:2020zkv}, since the problem becomes solvable thanks to the homogeneity and isotropy conditions. In the supernova context, currently, special efforts are being made to determine the impact of neutrino-neutrino interactions in core-collapse supernovae going beyond the mean-field approximation, either by exploring the role of many-body correlations also using concepts from quantum information and quantum simulations\cite{Balantekin:2023qvm}, or solving the neutrino quantum kinetic equations\cite{Nagakura:2025brr}.

A variety of flavor conversion phenomena have been uncovered in the last twenty years, due to the neutrino-neutrino neutral current interaction, shock wave effects and turbulence.
In particular, theoretical studies have uncovered that the $\nu\nu$ interaction, that makes the neutrino evolution equations non-linear\cite{Pantaleone:1992eq}, engender
slow modes first identified by Carlson et al\cite{Duan:2006an}, fast modes identified by Sawyer\cite{Sawyer:2015dsa} or collisional instabilities found by Johns\cite{Johns:2021qby}. While almost two decades of theoretical studies have paved our understanding of this complex many-body problem, still many aspects require further attention. 
 Besides the impact on future observations, an important aspect of such developments is the possibility
that flavor conversion impacts the supernova dynamics, a possibility long sought by theorists. Indeed fast conversions that were shown to occur in multi-dimensional supernova simulations\cite{Abbar:2018shq}, might help supernova explosions\cite{Ehring:2023abs}. 

\section{Future observations}
Supernovae are rare events. In our Galaxy, in the last thousand years, six events were observed namely SN1006, SN1572, SN1604 (SNIa) and SN1054, SN1181 and SN1667 (gravitational core-collapse supernovae). In the Local Group, in the last century, two more events were observed, i.e. SN1885 in Andromeda and SN1987A. There are many estimates of the gravitational core-collapse supernova rate including
$3.2^{+7.3} _{-2.6}$ events from historical supernovae in the Milky Way\cite{Adams:2013ana} and  the rate of $1.63 \pm 0.46$ events per century obtained by combining several available estimates (historical supernovae in the Milky Way, $^{26}$Al in our Galaxy, galactic neutron-stars, counting of massive stars at 1.5 kpc nearby the Sun,  supernovae in different morphological classes of host galaxies)\cite{Rozwadowska:2020nab}. 

If we observe a new core-collapse supernova event, we might detect a neutrino time signal going trom one-two days before the core-collapse\cite{Kato:2020lwd} up to 100 seconds after core-bounce. Indeed few MeV neutrinos are emitted by the cooling of the inner core, during the last Si-burning phase. Their detection would bring confirmation of stellar evolution theory, on the star progenitor and provide an early alert to astronomers and neutrino observers. For example, pre-SN neutrinos could be observed at $3 \sigma$ in KamLAND\cite{KamLAND:2015dbn} for 
a 25 $M_{sun}$ star at 690 kpc. On the other hand the late time emission, from 10 seconds to 100 seconds after bounce would inform us on the fate of the supernova,
the total radiated energy as neutrinos and non-standard cooling processes of the newly formed neutron star. For a supernova at 10 kpc, one could measure up to 250 $\bar{\nu}_e$
events during 50 seconds in Super-Kamiokande, 110 events over 40 seconds in DUNE and 10 $\nu_x$ ($x = \mu, \tau $) in JUNO\cite{Li:2020ujl}.

The Supernova Early Warning System\cite{SNEWS:2020tbu} (SNEWS) should alert astronomers if a supernova explosion takes place.
From the next galactic supernova if at 10 kpc, we will be able to precisely measure the neutrino time and energy signal and its different flavor components. 
Indeed we will detect from several hundred events in detectors like HALO-2 or KamLAND, 3 $10^3$ and 8 $10^3$ events in DUNE and JUNO respectively, $10^4$ in Super-Kamiokande, a few 10$^4$ in Hyper-Kamiokande, and 10$^6$ in IceCUBE. Moreover, from coherent neutrino-nucleus scattering in dark matter detectors we will observe for example 120 events in Xenon nT, about 700 events in DARWIN and 336 events in DarkSide-20 kt (see \cite{Volpe:2023met} and references therein).  

The detection of the different components of the 10 seconds neutrino time signal, namely the neutronization burst, the accretion phase and the cooling of the newly formed proto-neutron star will be crucial. For example, the early 20 ms $\nu_e$ signal of the neutronization burst should be affected only from the MSW effect, and offer the opportunity to search for new physics, such as non-radiative $\nu$ two-body decay or non-standard neutrino-matter interactions\cite{deGouvea:2019goq,Das:2017iuj}. Moreover it offers the possibility of a concurrent measurement of the gravitational waves emitted from core-bounce. Interestingly the detection of a supernova neutrino signal, also concomitantly with gravitational waves emitted due to the neutrino-driven convection, neutrino-heating in the gain layer and the SASI would offer a unique opportunity to confirm/refute the explosion mechanism since both
probes (neutrinos and gravitational waves) tell us about the evolution close to the stellar core. And, the measurement of the total neutrino luminosity with 10 $\%$ (3) $\%$ precision in Super-Kamiokande  (Hyper-Kamiokande) would provide tight constraints on the compactness (or mass-to-radius) relation of the newly formed neutron star\cite{GalloRosso:2017hbp}. 

Furthermore, ongoing studies based on Bayesian analyses show that we might discriminate among supernova models, and flavor mechanisms. 
In particular the first investigation\cite{Hyper-Kamiokande:2021frf} discussed the ability to disentangle five (one- or multidimensional) supernova models from different groups; whereas a following study\cite{Saez:2024ayk} focussed on 18 2D and 3D models from 9 $M_{\rm Sun}$ to 60 $M_{\rm Sun}$ progenitors.
Moreover the long-term emission of seven one-dimensional models (different progenitor and equation of state) was also explored\cite{Olsen:2022pkn}. These investigations came to the conclusion that the possibility to disentangle models, if a new supernova is observed, is in many cases excellent. 

\begin{figure}
\begin{center}
\includegraphics[scale=0.37]{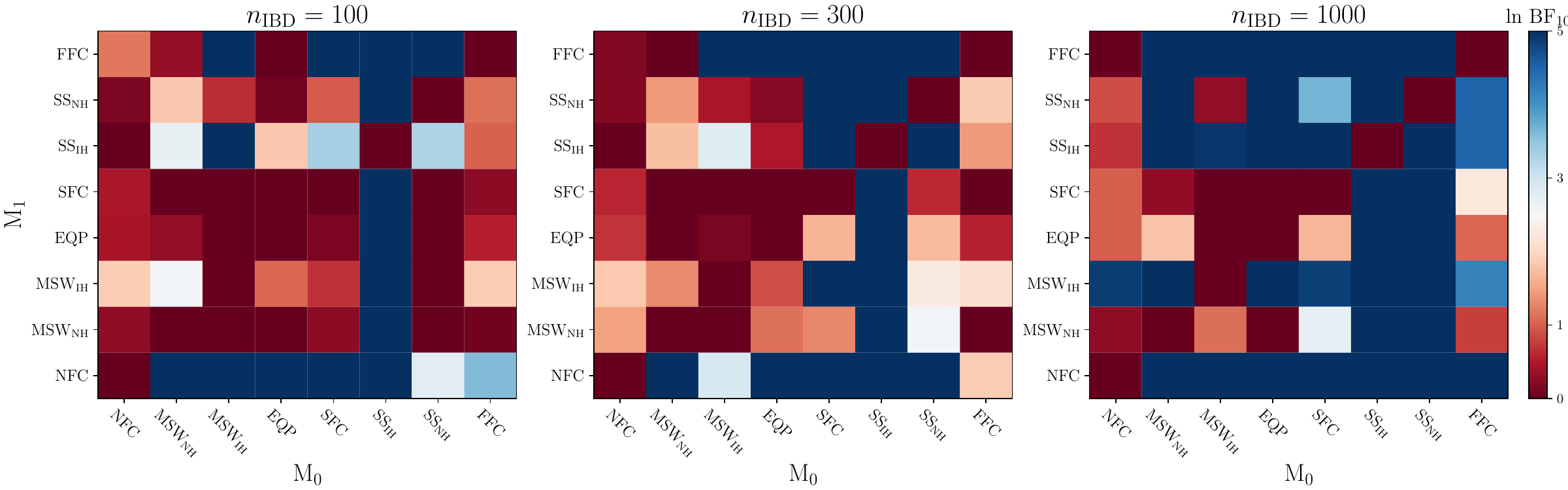}
\caption{Accretion phase of an exploding supernova and flavor evolution: heatmaps for the Bayes factor based on 100 (left), 300 (middle) to 1000 (right figure) inverse-beta decay events. The results assume neutrino flux parameters are not fixed and the location of the supernova unknown.. The acronysmes correspond to no-flavor conversion (NFC), the Mikheev-Smirnov-Wolfenstein effect for normal (MSW$_{\rm NH}$) or inverted (MSW$_{\rm IH}$) neutrino mass ordering, flavor equipartition (EQP), slow flavor conversion (SFC) modes, spectral swapping (SS$_{\rm NH}$, SS$_{\rm IH}$) fast flavor conversion (FFC). Note that if the logarithm of the Bayes factor is : i) (0-1), it is not significant; ii) (1–3), it indicates positive evidence; iii) (3–5): it reflects strong evidence; iv) it is $>$ 5, it demonstrates very strong evidence$^{70}$.}
\label{fig:flav}
\end{center}
\end{figure}

While previous studies only included the MSW effect, 
the first Bayesian analysis was also performed to address our ability to identify neutrino flavor mechanisms in a supernova\cite{Abbar:2024nhz}.
Figure \ref{fig:flav} presents heatmaps obtained considering inverse-beta decay events from 100 to 1000 (accretion phase, supernova distance unknown), assuming the neutrino flux parameters not to be fixed (but with some reasonable priors). One can see how the discriminating power evolves with the event number and the fact that 
even 1000 IBD events appear to be sufficient to disentangle some of the flavor evolution scenarios.  

Neutrinos from past supernovae constitute a neutrino background of all neutrino flavors, known as the diffuse supernova neutrino background (DSNB).
The DSNB encodes complementary information with respect to a single core-collapse supernova. Indeed the DSNB flux not only depends on the neutrino emission from
single supernovae that either turn into a neutron-star or into a black hole (failed supernovae), but also on the cosmological model and the evolving core-collapse supernova rate\cite{Volpe:2023met,Beacom:2010kk,Suliga:2022ica,Mathews:2019klh}.  
Important uncertainties on the DSNB predictions arise from the evolving core-collapse supernova rate, from the debated fraction of failed supernovae, the neutrino emission
from individual supernovae and also from the contributions from binaries. 

The combined analysis of Super-Kamiokande (SK) phases SK-I to SK-IV data (twenty years of data taking) in Super-Kamiokande found a 1.5 $\sigma$ excess over the background prediction (model dependent analysis). 
Interestingly, the combined sensitivity of SK-I to SK-IV data is on par with 4 DSNB predictions that have very different inputs\cite{Super-Kamiokande:2021jaq}.
Moreover, the analysis of the first results of Super-Kamiokande data with Gadolinium addition\cite{Harada:2024} (SK-VI to SK-VII) increased the statistical significance to 2.3 $\sigma$.

While the SK-Gd experiment has been taking data since 2020, upcoming experiments should have the discovery potential for the DNSB.
The Hyper-Kamiokande experiment\cite{Hyper-Kamiokande:2018ofw} that will have an active volume 8.3 larger than SK will start taking data in 2027, whereas 
the scintillator detector JUNO\cite{JUNO:2022lpc} (18 kt fiducial volume) should start in 2025 and the DUNE experiment\cite{DUNE:2020zfm} (40 kton liquid Argon) should start in a  in 2029-2032.  Depending on predictions, the expected number of events in ten years running typically ranges between 10 and 20 events in the SK-Gd experiment, 
about 10-40 in the JUNO experiment, 5-10 events in the DUNE experiment, 25-145  in the 
Hyper-K experiment. Such numbers should be considered as indicative of the most conservative to the most optimistic cases, while some of the 
predictions can yield an even larger number of events. 

The DSNB discovery will bring crucial information on the evolving core-collapse supernova rate, on the fraction of failed supernovae and on non-standard neutrino properties and particles. For example the DSNB will have a unique sensitivity\cite{Ando:2003ie,Fogli:2004gy,DeGouvea:2020ang,Ivanez-Ballesteros:2022szu,MacDonald:2024vtw,Roux:2024zsv} to non-radiative two-body decay in the window $\tau/m \in [10^9, 10^{11}]$ s/eV (Figure 1). An investigation\cite{Ivanez-Ballesteros:2022szu} bases on a $3\nu$ framework for the neutrino decay, including uncertainties from the evolving core-collapse supernova rate, showed that DSNB predictions with and without neutrino decay are degenerate for normal mass ordering and a strongly hierarchical mass pattern, whereas in inverted mass ordering neutrino decay could suppress the DSNB rate by a large factor. The first Bayesian analysis\cite{Roux:2024zsv} combining
inverse-beta decay rates in Hyper-Kamiokande, in JUNO and $\nu_e$-Argon scattering events in DUNE found that even combining the DSNB detection channels is not sufficient
to discriminate between the case with decay from the case without decay in the window $\tau/m \in [10^9, 10^{11}]$ s/eV. 

In the future a crucial input for the identification of the $r$-process sites will come from gravitational wave measurements
that will provide a precise binary neutron-star merger rate, and hopefully a new concomitant kilonova observation as well.
Future observations from the next (extra)galactic supernova will be crucial in particular to definitely assess the core-collapse supernova explosion mechanism, for our understanding 
of how neutrinos change flavor in dense environments and for the search for new physics. Indeed neutrinos from the next supernova will represent a unique laboratory to learn about flavor evolution in dense environments. In this respect recent Bayesian analyses appear as promising. Finally the discovery of the diffuse supernova background will open a novel and unique observational window in low energy neutrino astrophysics. 

\section*{Acknowledgments}

M. Cristina Volpe acknowledges  financial support from "CNRS Nucl\'eaire et Particules" through the Masterproject NUFRONT.

\end{document}